\documentclass[journal]{IEEEtran}

\ifCLASSINFOpdf

\else

\fi
\usepackage{soul}
\usepackage[utf8]{inputenc}
\usepackage{amsmath}
\usepackage{mathtools}
\usepackage{braket}
\usepackage{graphicx}
\usepackage{tikz}
\usepackage{quantikz}
\graphicspath{{./figures/}}
\usepackage{amsfonts}
\usepackage{comment}
\usepackage[ruled,vlined,linesnumbered]{algorithm2e}
\let\oldnl\nl
\newcommand{\nonl}{\renewcommand{\nl}{\let\nl\oldnl}}

\usepackage{csquotes}
\usepackage{hhline}
\usepackage{amssymb}
\usepackage{listings}
\usepackage{color}
\definecolor{codegreen}{rgb}{0,0.6,0}
\definecolor{codegray}{rgb}{0.5,0.5,0.5}
\definecolor{codepurple}{rgb}{0.58,0,0.82}
\definecolor{backcolour}{rgb}{0.95,0.95,0.92}

\usepackage{subcaption}
\usepackage{titlesec}

\usepackage{dcolumn}
\usepackage{tabularx}
\usepackage{hyperref}
\hypersetup{
    colorlinks=true,
    linktoc=all,
    linkcolor=black,
    citecolor=magenta,
}
\usepackage{longtable}
\usepackage{braket}
\newcolumntype{C}{>{\centering\arraybackslash}X}
\SetAlFnt{\small}
\usepackage[font=small]{caption}
\begin{document}
\hyphenation{op-tical net-works semi-conduc-tor}

\title{QNN-VRCS: A Quantum Neural Network for Vehicle Road Cooperation Systems}
\author{Nouhaila Innan\thanks{N.~Innan is with the Quantum Physics and Magnetism Team, LPMC, Faculty of Sciences Ben M'sick, Hassan II University of Casablanca, Morocco; And with the eBRAIN Lab, Division of Engineering, New York University Abu Dhabi (NYUAD), and the Center for Quantum and Topological Systems (CQTS), NYUAD Research Institute, NYUAD, Abu Dhabi, UAE.
 e-mail: (nouhaila.innan@nyu.edu).}, Bikash~K.~Behera\thanks{B.~K. Behera is with the Bikash's Quantum (OPC) Pvt. Ltd., Mohanpur, WB, 741246 India, e-mail: (bikas.riki@gmail.com).}, Saif Al-Kuwari\thanks{Saif Al-Kuwari is with the Qatar Center for Quantum Computing, College of Science and Engineering, Hamad Bin Khalifa University, Qatar Foundation, Doha, Qatar. e-mail: (smalkuwari@hbku.edu.qa).}, and Ahmed~Farouk\thanks{A.~Farouk is with the Department of Computer Science, Faculty of Computers and Artificial Intelligence, South Valley University, Hurghada, Egypt. e-mail:(ahmed.farouk@sci.svu.edu.eg).}}%

\maketitle

\begin{abstract}
The escalating complexity of urban transportation systems, exacerbated by factors such as traffic congestion, diverse transportation modalities, and shifting commuter preferences, necessitates the development of more sophisticated analytical frameworks. Traditional computational approaches often struggle with the voluminous datasets generated by real-time sensor networks, and they generally lack the precision needed for accurate traffic prediction and efficient system optimization. This research integrates quantum computing techniques to enhance Vehicle Road Cooperation Systems (VRCS). By leveraging quantum algorithms, specifically $UU^{\dagger}$ and variational $UU^{\dagger}$, in conjunction with quantum image encoding methods such as Flexible Representation of Quantum Images (FRQI) and Novel Enhanced Quantum Representation (NEQR), we propose an optimized Quantum Neural Network (QNN). This QNN features adjustments in its entangled layer structure and training duration to better handle the complexities of traffic data processing. Empirical evaluations on two traffic datasets show that our model achieves superior classification accuracies of 97.42\% and 84.08\% and demonstrates remarkable robustness in various noise conditions. This study underscores the potential of quantum-enhanced 6G solutions in streamlining complex transportation systems, highlighting the pivotal role of quantum technologies in advancing intelligent transportation solutions.
\end{abstract}


\IEEEpeerreviewmaketitle

\section{Introduction}\label{QVP:Sec1}

Vehicle road cooperation systems (VRCS) transform urban mobility by enabling dynamic interactions between vehicles and road infrastructure, significantly improving traffic efficiency, safety, and reducing environmental impacts \cite{cascetta1993modelling, olia2016assessing}. As an illustration of smart city technology, VRCS adapts to the challenges posed by the 1.5 billion cars worldwide, underscoring the limitations of traditional traffic management and highlighting the need for innovative solutions \cite{erhardt2019transportation}.
However, VRCS's reliance on heuristic methods, which adhere to predefined rules, poses significant challenges in managing the unpredictable dynamics of road traffic \cite{pisinger2007general}. This inefficiency often leads to suboptimal decisions during complex traffic scenarios, underscoring the necessity for more sophisticated and adaptable technologies. 
The integration of artificial intelligence (AI) and the 6G promises to revolutionize VRCS by enhancing predictive analytics and decision-making capabilities \cite{mishra2022leveraging, zhang2019mobile}. Despite these advancements, deploying such technologies within VRCS requires substantial computational resources and challenges ensuring reliability across diverse transportation scenarios \cite{siegel2017survey}.
In parallel, the development of fully autonomous vehicles, supported by advancements in driver-assistance systems, relies on sophisticated visual processing for navigation and obstacle detection \cite{bengler2014three}. Quantum computing (QC) is a potential solution to enhance the computational efficiency and accuracy required for VRCS and autonomous vehicle technologies \cite{preskill2023quantum, salek2023hybrid}. By integrating QC with AI through quantum machine learning (QML), particularly quantum neural networks (QNNs), new avenues are opened for overcoming the computational constraints and robustness challenges faced by current models \cite{biamonte2017quantum}.
The adaptability and efficiency of QNNs make them particularly suitable for addressing the complex challenges within VRCS. This paper explores a novel approach that combines QC with advanced AI techniques in VRCS, focusing on the potential of QML models, especially QNNs, to tackle the multifaceted issues of traffic management. The integration aims to contribute to the development of fully autonomous driving systems and the realization of optimized VRCS, indicating a new era of intelligent mobility.

\begin{figure}
    \centering
    \includegraphics[width=\linewidth]{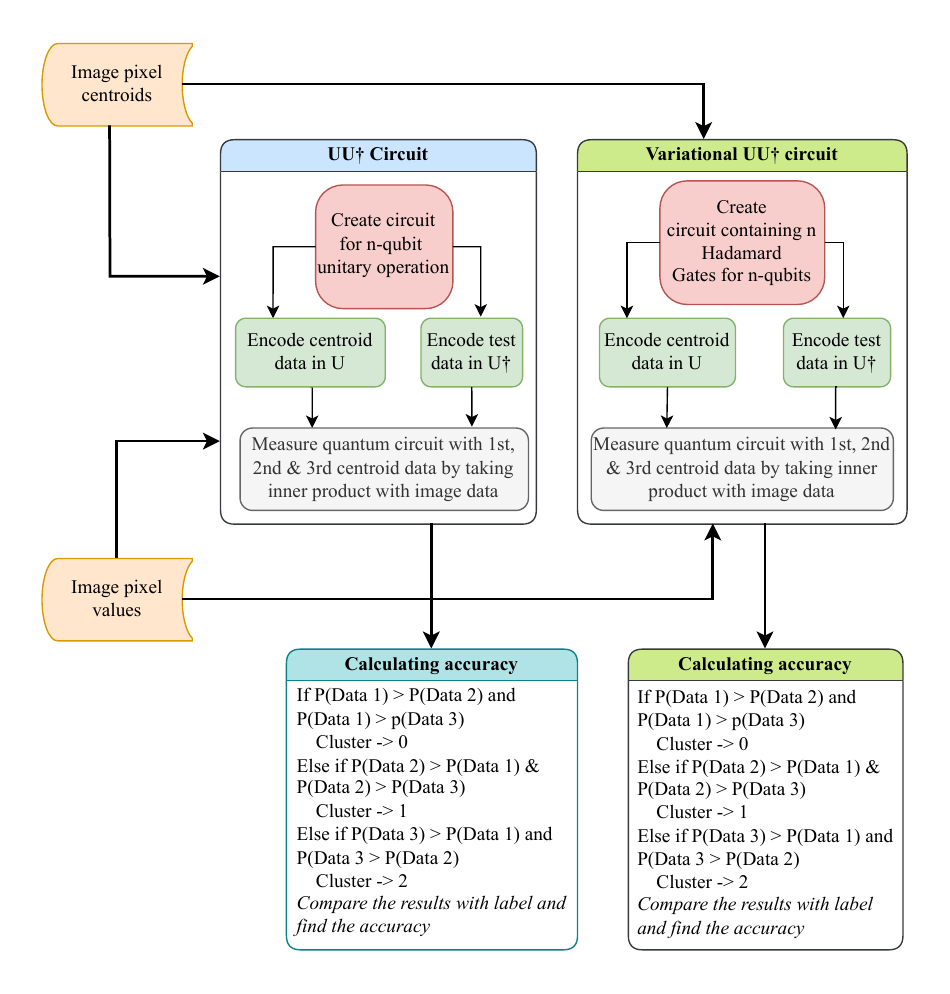}
    \caption{Schematic diagram showing the working of $UU^{\dagger}$, and variational $UU^{\dagger}$ algorithms.}
    \label{fig:1a}
\end{figure}

\begin{figure*}
    \centering
    \includegraphics[width=\linewidth]{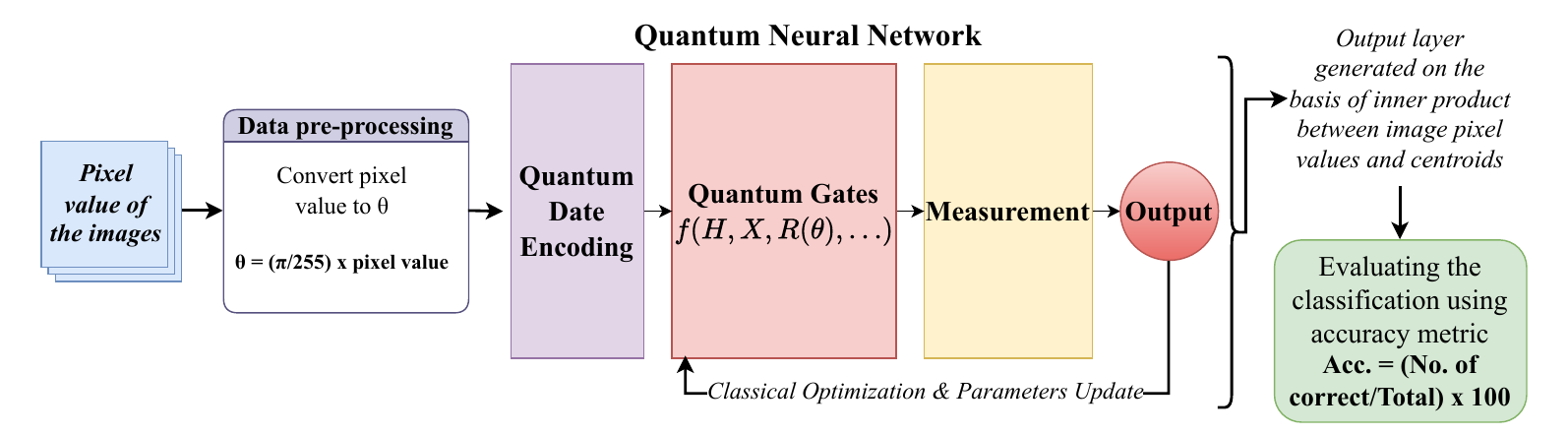}
    \caption{Schematic diagram showing the working of QNN algorithm.}
    \label{fig:1b}
\end{figure*}

\subsection{Related Works}

Quantum computing has previously been explored for traffic prediction and control, using hybrid classical-quantum models and specific quantum algorithms aimed at enhancing traffic management \cite{kuros2022traffic,qt_tt}. This research employs a QNN approach for classifying traffic light images to evaluate their efficacy in automated vehicle vision systems. While the capability of QNNs to consistently outperform classical networks remains under investigation, early studies suggest that QNNs can provide more accurate representations of quantum states than classical models \cite{schuld2017, abbas2021}.
QML may offer polynomial speed improvements over classical algorithms for complex tasks, despite the exponential speedups being more characteristic of specific algorithms like Shor's \cite{schuld2017}. This potential makes QML particularly interesting for intricate applications such as traffic light image classification.
In this context, we utilize QNNs, alongside $UU^{\dagger}$ and variational $UU^{\dagger}$ methods, leveraging quantum image representation techniques like flexible representation of quantum images (FRQI) and novel enhanced quantum representation (NEQR) to encode images \cite{qt_frqi, qt_neqr}. These methods are tailored for quantum image analysis, where FRQI and NEQR convert pixel values into quantum states for processing. The QSobel algorithm, an advancement over conventional edge detection methods, demonstrates significant computational efficiency in edge extraction from quantum images, using only one qubit per pixel to store grayscale values, thereby simplifying the image processing pipeline \cite{qt_qsobel}.
Further, our study extends to using these quantum computing (QC) techniques for data clustering and classification, incorporating them into classical-quantum hybrid frameworks for practical applications. This includes real-world implementations like the live QC-based navigation system introduced at the web summit in Lisbon and the development of an intelligent transportation system (ITS) using D-Wave's quantum technology, showcasing the applicability of both universal and specialized quantum computers in transportation \cite{yarkoni2020quantum, wang2021shaping}. Here, we propose the $UU^{\dagger}$, variational $UU^{\dagger}$, and QNN algorithms, whose schematic diagrams are shown in Figs. \ref{fig:1a} and \ref{fig:1b}.

\subsection{Novelty and Contribution}

\begin{itemize}
\item[1)] Development and implementation of an innovative QNN designed for VRCS.
\item[2)] Application of variational quantum algorithms for classification tasks and quantum methods for data encoding, showcasing the adaptability and efficiency of QC techniques in complex classification challenges.
\item[3)] A comparative analysis between classical and quantum ML approaches across two datasets (Carla and cropped Lisa), highlighting the superior accuracy of QNNs.
\item[4)] Evaluation of the QNN's robustness against six noise models, assessing its resilience in varied and challenging conditions.
\end{itemize}

\subsection{Organization}

\textit{Structure:} The rest of the paper is organized as follows: Sec. \ref{sec2} defines the problem and overall framework, and the methodology of the algorithms. Sec. \ref{sec4} presents the experimental setup, two datasets, preprocessing, and results for both the datasets, noisy simulations and discussion. Finally, Sec. \ref{sec6} concludes the paper, summarizing the findings and suggesting future research directions.

\section{Methodology}

\subsection{Problem Overview\label{sec2}}

The objective is to develop a QNN model $f_{\theta}(x)$ that classifies traffic light states as part of an optimization problem. We employ the dataset $\mathcal{D} = \{(x_i, y_i)\}_{i=1}^{N}$, consisting of $N$ pairs of input images $x_i$ from the Carla and cropped Lisa datasets, along with their corresponding labels $y_i$. The goal is to find the optimal parameters $\theta$ for the QNN model, which aims to outperform both classical and other quantum models in traffic light state classification. The input images $x_i$ are converted into quantum states using angle encoding, adapted for the dimensionality of $2 \times 2$. Our approach is to minimize a general loss function $L(\theta)$, which measures the discrepancy between the model's predictions $f_{\theta}(x_i)$ and the actual labels $y_i$. The specific form of $L(\theta)$ will be detailed in subsequent sections. This formulation lays the groundwork for employing quantum optimization techniques to refine $\theta$, thereby enhancing the model's classification accuracy.

\subsection{Encoding Methods}

\subsubsection{Flexible Representation of Quantum Images (FRQI)}

FRQI is one of the encoding algorithms for image processing. The quantum state containing the image is represented by,

\begin{equation}
    \ket{I(\theta)}= \frac{1}{2^{n}} \sum_{i=0}^{2^{2n}-1} (\cos\theta_{i}\ket{0} +\sin\theta_{i}\ket{1}) \otimes \ket{i}
\end{equation}

where, $\theta_{i} \in [0, \pi/2], i=0,1,....,2^{2n}-1$. Here $I(\theta)$ is a normalized state and consists of two parts:

\begin{itemize}
    \item Color information encoding: $\cos{\theta_{i}}\ket{0}+\sin{\theta_{i}}\ket{1}$
    \item Associated pixel position encoding: $\ket{i}$
\end{itemize}

In our data, first, all the images are resized into $2 \times 2$ and $4 \times 4$ dimensional forms. Then, the resized images are converted into their array representation, where the centroid values of rows and columns are stored in matrix form. These centroid values of the matrices are then converted into angles ranging from 0 to $\pi$ using the formula 
\begin{equation}
  \theta = \text{Pixel Value} \times \frac{\pi}{255}
\end{equation}

and then substituted in the value of $\theta$ in the circuit. This completes our FRQI encoding method of images. For a $2\times 2$ image, the FRQI circuit is given in Fig. \ref{Fig3} (a). 

\subsubsection{Novel Enhanced Quantum Representation (NEQR)}

NEQR \cite{bib_NEQR} uses a series of qubits to represent the intensity value of pixels presented by a set of $CNOT$ gates (Fig. \ref{Fig3} (b)). 
For representing two-dimensional images, we define the position of the image by its row and column, $Y$, $X$, respectively. For a single bit, $0$ represents black, and $1$ represents white. Grayscale images that have intensity values ranging from $0$ to $255$ are represented by $8$ bits, while color images are represented by $24$ bits broken up into $3$ groups of $8$ bits, where each group of $8$ bits represents the pixel color of red, green and blue. 

Thus, each pixel value can be represented as per the following expression:

\begin{eqnarray}
f(Y, X)&=&{C_{Y X}}^{0}, {C_{Y X}}^{1}, \ldots C_{Y X}^{q-2}, C_{Y X}^{q-1} \in[0,1].
\end{eqnarray}

Here $f(Y, X)\in\left[0,2^{q-1}\right]$ and $C$ is a binary representation of the grayscale intensity value. The general expression to represent an image for a \textit{\textbf{$2^n \times 2^n$}} dimension $|I\rangle$ is given as:

\begin{eqnarray}
|I\rangle&=&\dfrac{1}{2^{n}} \sum_{Y=0}^{2^{2 n-1}} \sum_{X=0}^{2^{2 n-1}}|f(Y, X)\rangle|Y X\rangle,\\
&=& \dfrac{1}{2^{n}} \sum_{Y=0}^{2^{2 n-1}} \sum_{X=0}^{2^{2 n-1}}\left|\otimes_{i=0}^{q-1}\right\rangle\left|C_{Y X}^{i}\right\rangle|Y X\rangle.
\end{eqnarray}

\subsection{Classification Methods}

\begin{algorithm}[!t]
\DontPrintSemicolon
\nonl \textbf{Input} Image datasets\;
\nonl \textbf{Output} Classification accuracy\;
Assemble the quantum circuit utilizing an $n$-qubit unitary operation $U$ and an $n$-qubit $U^{\dagger}$ operation \;
Embed the centroid information into the $U$ operator, resulting in the states $\ket{\psi_{1}}$, $\ket{\psi_{2}}$, and $\ket{\psi_{3}}$ corresponding to clusters $0$, $1$, and $2$ respectively, and incorporate the test data into the $U^{\dagger}$ operator, producing $\ket{\phi_{i}}$. \;
Execute measurements on the quantum circuit starting with the first, then the second, and finally the third centroid data.\;
{\textbf{If} ${\langle \psi_{1} | \phi_{i} \rangle > \langle \psi_{2} | \phi_{i} \rangle}$ and ${\langle \psi_{1} | \phi_{i} \rangle > \langle \psi_{3} | \phi_{i} \rangle}$ \textbf{then} classify the respective test data as $\text{cluster - 0}$}\;
{\textbf{ElseIf} ${\langle \psi_{2} | \phi_{i} \rangle > \langle \psi_{1} | \phi_{i} \rangle}$ and ${\langle \psi_{2} | \phi_{i} \rangle > \langle \psi_{3} | \phi_{i} \rangle}$ \textbf{then} classify the respective test data as $\text{cluster - 1}$}\;
{\textbf{Else} classify the respective test data as $\text{cluster - 2}$}\;
{\textbf{EndIf}}\;
{Validate the classifications against actual test data to determine if the data corresponds to red, yellow, or green images}\;
If $c_{1}$, $c_{2}$, and $c_{3}$ represent the classification accuracies for red, yellow, and green images respectively, and $N_{1}$, $N_{2}$, and $N_{3}$ denote the probabilities of these image categories respectively (Number of specific images/ Total number of images), then compute the overall accuracy using the formula:\;
\begin{center}
$c_{1} * N_{1} + c_{2} * N_{2} + c_{3} * N_{3}$
\end{center}
\caption{$UU^\dagger$ 
}
\label{algo1}
\end{algorithm}

\begin{algorithm}[t]
\nonl \textbf{Input} Image data collection \;
\nonl \textbf{Output} Classification accuracy \;
\For{$m \in \mathbb{N}$}{
    {Set up the circuit with two Hadamard gates followed by an $n$-qubit unitary operation $U$, and its conjugate transpose $U^{\dagger}$, ending with two more Hadamard gates.} \;
    {Embed the centroid information into the $U$ operator to produce the states $\ket{\psi_{1}}$, $\ket{\psi_{2}}$, and $\ket{\psi_{3}}$, corresponding to clusters 0, 1, and 2 respectively, and encode the test data in the $U^{\dagger}$ operator to obtain $\ket{\phi_{i}}$.} \;
    {Execute measurements on the quantum circuit starting with the data from the first centroid, followed by the second and third.} \;
    {\textbf{If} ${\langle \psi_{1} | \phi_{i} \rangle > \langle \psi_{2} | \phi_{i} \rangle}$ and ${\langle \psi_{1} | \phi_{i} \rangle > \langle \psi_{3} | \phi_{i} \rangle}$ \textbf{then} classify the respective test data as $\text{cluster - 0}$} \;
    {\textbf{ElseIf} ${\langle \psi_{2} | \phi_{i} \rangle > \langle \psi_{1} | \phi_{i} \rangle}$ and ${\langle \psi_{2} | \phi_{i} \rangle > \langle \psi_{3} | \phi_{i} \rangle}$ \textbf{then} classify the respective test data as $\text{cluster - 1}$} \;
    {\textbf{Else} classify the respective test data as $\text{cluster - 2}$} \;
    \textbf{EndIf} \;
    {Evaluate the classification results against the actual test dataset to see if the data corresponds to red, yellow, or green images} \;
    {If $c_{1}$, $c_{2}$, and $c_{3}$ are the accuracies for red, yellow, and green images respectively, and $N_{1}$, $N_{2}$, and $N_{3}$ denote the probabilities of these colors respectively (number of specific images/total number of images), then compute the overall accuracy using:} \;
    \begin{center}
    $c_{1} * N_{1} + c_{2} * N_{2} + c_{3} * N_{3}$
    \end{center}}
\caption{Variational $UU^\dagger$ 
for $m$ layers}
\label{algo2}
\end{algorithm}

\subsubsection{ $UU^{\dagger}$ Algorithm \label{BbVP}}

\begin{figure}[]
\centering
\begin{subfigure}{\linewidth}
\includegraphics[width=\linewidth]{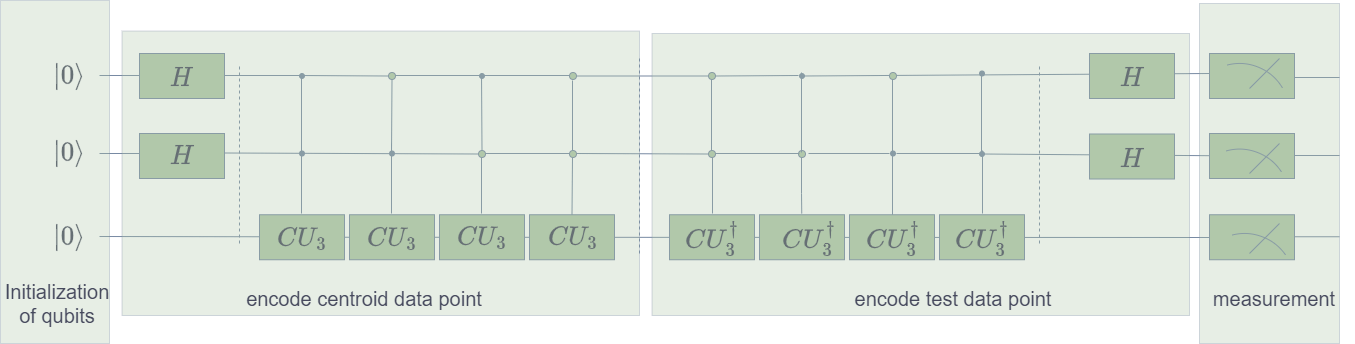}
    \caption{}
\end{subfigure}\hfill
\begin{subfigure}{\linewidth}
    \includegraphics[width=\linewidth]{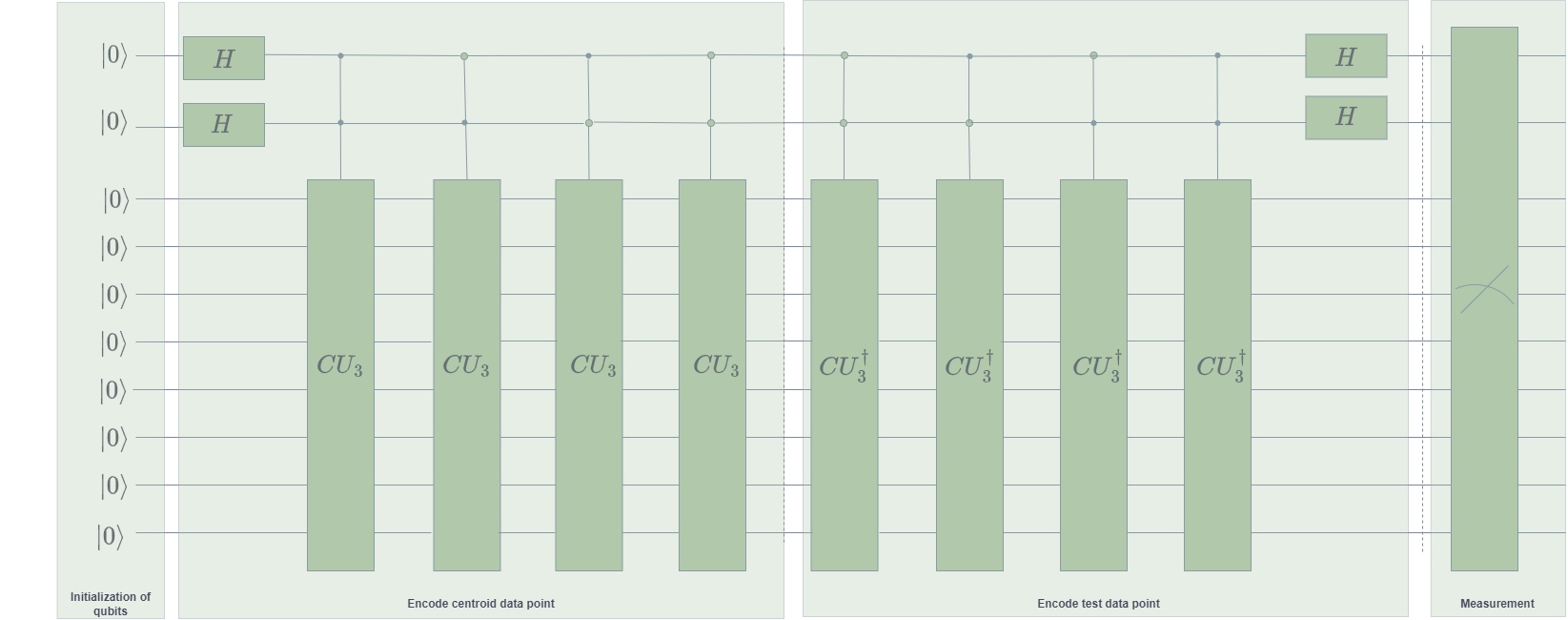}
\caption{}
\end{subfigure}
\begin{subfigure}{\linewidth}
\centering
    \includegraphics[width=0.9\linewidth]{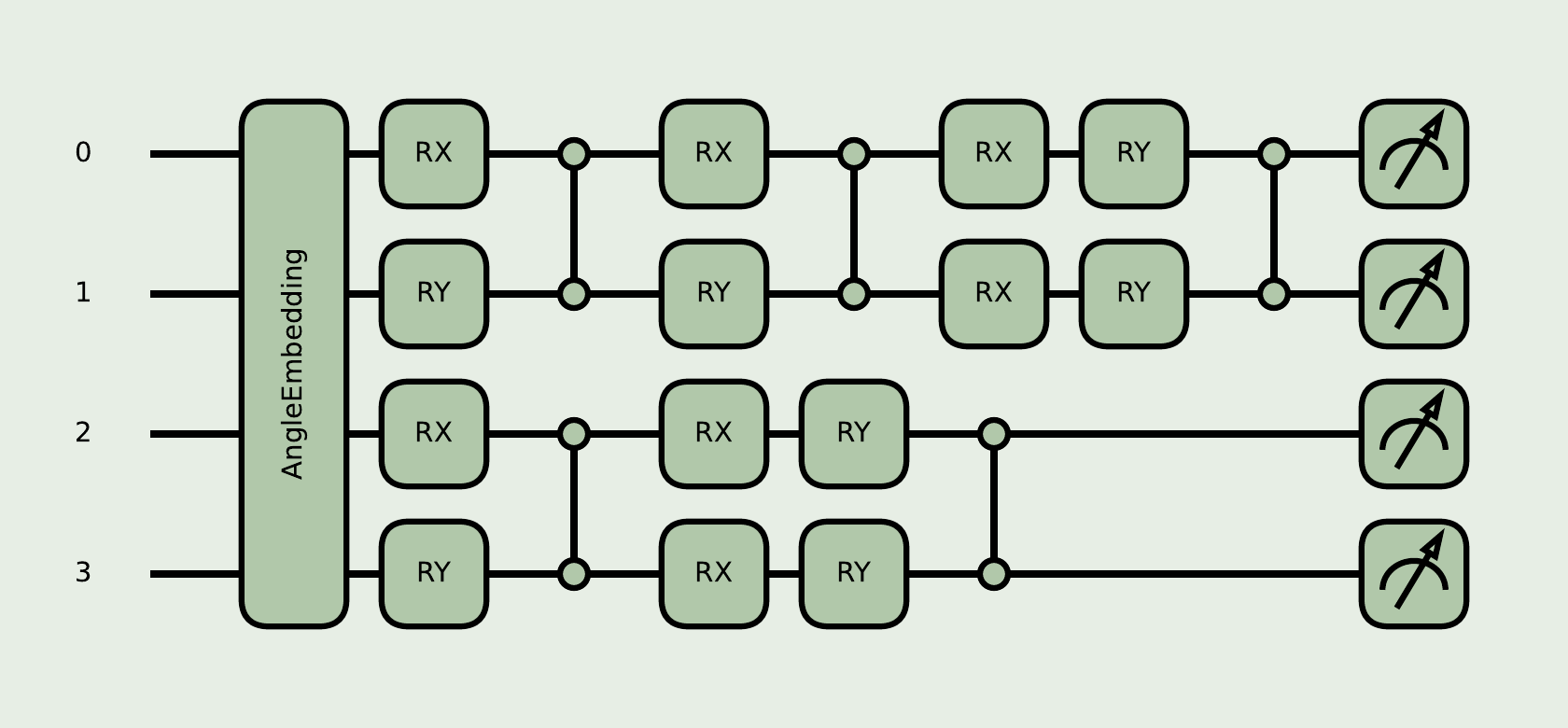}
\caption{}
\end{subfigure}
\caption{(a) Circuit for $UU^{\dagger}$ methodology using FRQI encoding, (b) Circuit for $UU^{\dagger}$ methodology using NEQR encoding, and (c) Circuit for QNN.}
\label{Fig3}
\end{figure}

In this classification algorithm, we first choose a centroid of each image before building the circuit. 
Each qubit in the quantum circuit is manipulated by controlled-$U3$, anti-controlled-$U3$, and $U3$ gates, with the centroid of every image stored in each gate. The $U$ gates record all $\theta$ features translated from angle $0$ to $\pi$, while the dagger gates encode the red, green, and yellow centroid used in image generation. The probability of $\ket{00}$ is then determined.
Similarly, the process is repeated for the second and third centroids of each image. 
If the probability of the state $\ket{00}$ is higher for the first centroid than for the second, it is in the red signal cluster. The same is true for the green and yellow signal clusters. Afterwards, the method's accuracy is determined.
Here, the states $\ket{\psi_{1}}$, $\ket{\psi_{2}}$ and $\ket{\psi_{3}}$ are prepared using the centroids of red, green, and yellow signal respectively. Unknown data value is stored in U gates and $\ket{\phi_{i}}$ state is prepared. Now, measurement is performed, followed by calculating the inner product between $\ket{\psi_{1}}$ and $\ket{\phi_{i}}$.
Also, the inner products between $\ket{\psi_{2}}$ ,$\ket{\phi_{i}}$ and $\ket{\psi_{3}}$, $\ket{\phi_{i}}$ are calculated. If ${\langle \psi_{1} | \phi_{i} \rangle > \langle \psi_{2} | \phi_{i} \rangle}$ and ${\langle \psi_{1} | \phi_{i} \rangle > \langle \psi_{3} | \phi_{i} \rangle}$ then, the data point belongs to red and likewise we will get for green and yellow. The example circuits for both FRQI and NEQR encoding using the analytical clustering method are presented in Fig. \ref{Fig3}. The quantum states for the centroid and test data points are given as,

\begin{eqnarray}
U_1\ket{0}^{\otimes n}=\ket{\psi_1}, \ U_2\ket{0}^{\otimes n}=\ket{\psi_2}
\end{eqnarray}

The inner product between the centroid and test data points is determined as follows \cite{qt_Deshmukh},
\begin{eqnarray}
    \braket{\psi_1|\psi_2}=^{n \otimes}\bra{0}U_1^{\dagger}U_2\ket{0}^{\otimes n}=\sqrt{P_0}
\end{eqnarray}

\subsubsection{Variational $UU^{\dagger}$ Algorithm} \label{VRP}

The variational $UU^\dagger$ technique of classification uses the exact same steps as the $UU^\dagger$ method,
but includes a modification where the Hadamard gate is positioned both before the measurement phase and subsequent to the $\ket{0}$ state initialization. 
and the circuit consists of equally repeating layers of $U$ and $U^\dagger$ gates. 
We repeat the $U$ operation $n$ times, followed by the $U^\dagger$ $n$ times, to increase the layering in the circuit. After that, we take measurements and determine the likelihood of $\ket{00}$. If the probability ($\ket{00}$) for a data point relative to the first centroid (red) exceeds those for the second (green) and third centroids (yellow) 
(${\langle \psi_{1} | \phi_{i} \rangle > \langle \psi_{2} | \phi_{i} \rangle}$ and ${\langle \psi_{1} | \phi_{i} \rangle > \langle \psi_{3} | \phi_{i} \rangle}$), then it is classified within the red cluster, and similarly for other clusters. Accuracy assessments employ the previously established methodologies. 

\subsubsection{Quantum Neural Network Algorithm}

Variational Quantum Circuits (VQCs) are crucial in QC, supporting various numerical tasks, including classification, optimization, and predictions. These circuits are fundamental to the variational quantum algorithm (VQA), an innovative hybrid approach that integrates classical and QC. In this paradigm, classical computers play a crucial role by optimizing the parameters of the VQC, thus enhancing the overall computational efficiency.
Within QNNs, the variational parameters are intricately encoded within the rotation angles of specific quantum gates, forming parameterized quantum circuits. The rotation gates most commonly employed in these settings are the single-qubit rotation gates $R_{x}(\theta)$, $R_{y}(\theta)$, and $R_{z}(\theta)$, in addition to controlled-Not gates and controlled-$Z$ gates. These gates are pivotal in manipulating quantum states and are central to the operational execution of quantum algorithms within VQC and VQA frameworks. This intricate interplay of quantum gates facilitates a robust platform for exploring and using quantum phenomena in practical computational applications.
The variational circuit incorporates three principal elements: a parameterized quantum circuit $U(x,\theta)$ \cite{que_var}, the resultant quantum output $y_{\text{predicted}i}$, and an iterative method for updating the parameters $\theta$. The data $x$ is pre-processed on a classical device to determine the input quantum state. Quantum hardware computes $U(x; \theta)$ using initialized parameters $\theta$ on a quantum state $\ket{x}$. After multiple $U(x; \theta)$ executions, the classical part analyzes measurements to generate a prediction $y_{\text{predicted}_i}$. The parameters are updated, and this cyclic process continues between classical and quantum devices. 
\begin{eqnarray}
R_y(\theta) =
\begin{bmatrix}
\cos{\theta} & - \sin{\theta}\\
\sin{\theta} & \cos{\theta} 
\end{bmatrix},
H = \frac{1}{\sqrt{2}} \begin{bmatrix}
   1 & 1 \\
   1 & -1 \\
\end{bmatrix},
\end{eqnarray}
\begin{eqnarray}
U_3 = \begin{bmatrix}
   \cos(\theta/2) & -e^{i\lambda} \sin(\theta/2) \\
   e^{i\phi} \sin(\theta/2) & e^{i(\phi+\lambda)} \cos(\theta/2)
\end{bmatrix},
\end{eqnarray}
\begin{eqnarray} 
CNOT = 
\begin{bmatrix}
I & O\\
O & X
\end{bmatrix},
CZ = 
\begin{bmatrix}
I & O\\
O & Z
\end{bmatrix},
CU_3= \begin{bmatrix}
   I & O \\
   O & U_3
\end{bmatrix},
\end{eqnarray}
where X and Z are Pauli X and Z gates respectively.
The loss function determines how far off from the optimum response we are, and its expression is described in Eq.\eqref{ref2}
\begin{equation}
\text{loss function} = (y_{\text{predicted}_i} - y_{\text{actual}_i} )^{2},
\label{ref2}
\end{equation}

where $y_{\text{actual}_i}$ represents the actual value and $y_{\text{predicted}_i}$ the forecasted value. The loss function is computed for the entire training dataset, with its average referred to as the cost function ($C.F.$).

\begin{equation}
C.F. = \frac{1}{n} \sum_{i=1}^{n} (y_{\text{predicted}_i} - y_{\text{actual}_i})^{2}.
\end{equation}


The rule to update weights is given by 
\begin{equation}
\theta_{i} = \theta_{i} - \eta \frac{\partial C.F.}{\partial \theta_{i}}
\end{equation}

The chain rule can be used in this situation to find the partial derivative. The entire procedure is fully described by the algorithm \ref{algo3}.

\begin{algorithm}[t]
\nonl \textbf{Input} Imaging data \;
\nonl \textbf{Output} Classification accuracy\; 
    {Do the pre-processing of the images}\;
    {Convert the pixels of the images in to $\theta$ value from 0 to $\pi$ using the relation}\;
    \nonl $$\frac{\pi}{255} \times \text{pixel value}$$
    \For{$m \in \mathbb{N}$}{
    {Classify the images classically into red, yellow, and green based on the maximum inner product of the images with their centroids}\;
    {Initialize all the qubits in $|0\rangle^{\otimes n}$ state} \;
    {Encode the pixel values using the rotational $X$ operators}\;
    {Apply entangled layer/layers using rotational $Y$ and CNOT gates}\;
    {
    Measure the circuit, compute loss, and update weights}\;}
    {Compare the quantum classification with the classical one}\;
    {Calculate the accuracy of the quantum classification}
\caption{QNN}
\label{algo3}
\end{algorithm}


\section{Experimental Results\label{sec4}}

\subsection{Settings}

This study is structured around the following research questions:
\begin{enumerate}
\item How does the QNN model's performance in vehicle transportation and light traffic section classification compare with that of established classical and quantum models?
\item How does varying the number of layers in the QNN model affect its accuracy and loss?
\item How stable is the QNN model under different quantum noise conditions, especially in tasks related to traffic classification?
\end{enumerate}
An extensive evaluation is conducted to address (1), contrasting the QNN model with CNN. This evaluation uses distinct datasets specifically for vehicle transportation and light traffic section classification, emphasising assessing the models using loss metrics and accuracy.
A detailed comparative analysis is performed between the QNN model and other QML models such as $UU^{\dagger}$ and variational $UU^{\dagger}$. 
To answer (2), the impact of varying the number of layers in the QNN model on its accuracy and loss metrics is directly evaluated.
and to answer (3), the resilience of the QNN model is tested against a series of six different quantum noise scenarios. This aspect of the research seeks to assess the model's robustness and accuracy in environments with varying levels of quantum noise.

\subsection{Datasets}\label{QVP:Sec2}

The Carla traffic lights dataset contains images of traffic lights that are followed while driving vehicles \cite{Kaggledataset1}. The dataset contains $1909$ images, among which $664$ images are red, $586$ images are yellow, and $659$ images are green. 
The Cropped Lisa traffic light dataset consists of $36,534$ cropped images sourced from the Lisa traffic light dataset \cite{Kaggledataset2}. These images are distributed among seven directories for both training and testing purposes, while
our classification task is concentrated solely on the `stop,' `warning,' and `go' directories.
With these two datasets, the models can be considered more comprehensive, as the simulated data provided by Carla are in a controlled environment for the performance of traffic light classification in predictable situations, while the real-world data provided by Lisa expose the models to more complex and highly diverse scenarios, including variable lighting conditions and driving environments.

\subsection{Preprocessing}

The preprocessing pipeline begins by loading a pre-trained traffic light detection model from TensorFlow, trained on a large object detection dataset. This model identifies traffic light objects within each image. All images are resized to both $2\times2$ and $4\times 4$ pixels depending on the model being used. After resizing, the pixel values are normalized by scaling between 0 and 1. The object detection model is applied to the preprocessed images to identify the relevant traffic light objects. Bounding boxes are extracted for objects that satisfy the following predefined criteria: the object must belong to $C$, and the detection score also needs to be higher than $T$, where, $I$ denotes the imaging dataset, $T$ represents the score threshold, and $C$ denotes the target class of traffic light objects.

\subsection{Hyperparameters}

In our implementation, the number of qubits is determined based on the image size and the encoding method. The QNN uses angle encoding and requires 4 qubits for a $2 \times 2$ input. The $UU^\dagger$ and variational $UU^\dagger$ models are implemented using the NEQR and FRQI encoding methods, respectively. NEQR encoding requires 10 qubits for a $2 \times 2$ input and 12 qubits for a $4 \times 4$ input, while FRQI encoding requires fewer qubits, namely 3 qubits for a $2 \times 2$ input and 5 qubits for a $4 \times 4$ input. For the depth of quantum circuits, We vary the number of layers across 1, 5, 10, 15, and 20, focusing on 20 layers as the optimal choice.
In parallel, the CNN architecture consists of Conv2D layers, MaxPooling2D layers, BatchNormalization, Dropout, and Dense layers, employing ReLU activation in the hidden layers and softmax for the output layer.
Both QNN and CNN models undergo training for 20 epochs with a consistent batch size of 32 across all datasets. The data is split into 80\% for training and 20\% for testing to ensure a balanced evaluation of model performance. The quantum and classical models utilize the Adam optimizer. This learning rate can be varied between 0.001 and 0.1, and it has been set to its optimum value at 0.001 for the quantum models and 0.01 for the CNN. For the CNN model, the TensorFlow/Keras is employed, whereas for quantum models, we use the Qiskit and PennyLane, with IBM Quantum providing noise simulations for the QNN model.

\subsection{Noise Models}

Noise models are mathematical frameworks used to describe the influence of noise and errors on quantum systems. A key approach for articulating these models employs Kraus operators, which constitute a set capable of representing any quantum error channel. These channels process a quantum state as input and yield another quantum state as output, incorporating a certain probability of error occurrence. The QNN algorithm is evaluated against six specific noise models: bitflip, phaseflip, bitphaseflip, depolarizing, amplitude damping, and phase damping \cite{qt_sks}.
\begin{figure}[htpb]
    \centering
\begin{subfigure}{\linewidth}
    \includegraphics[width=0.8\linewidth]{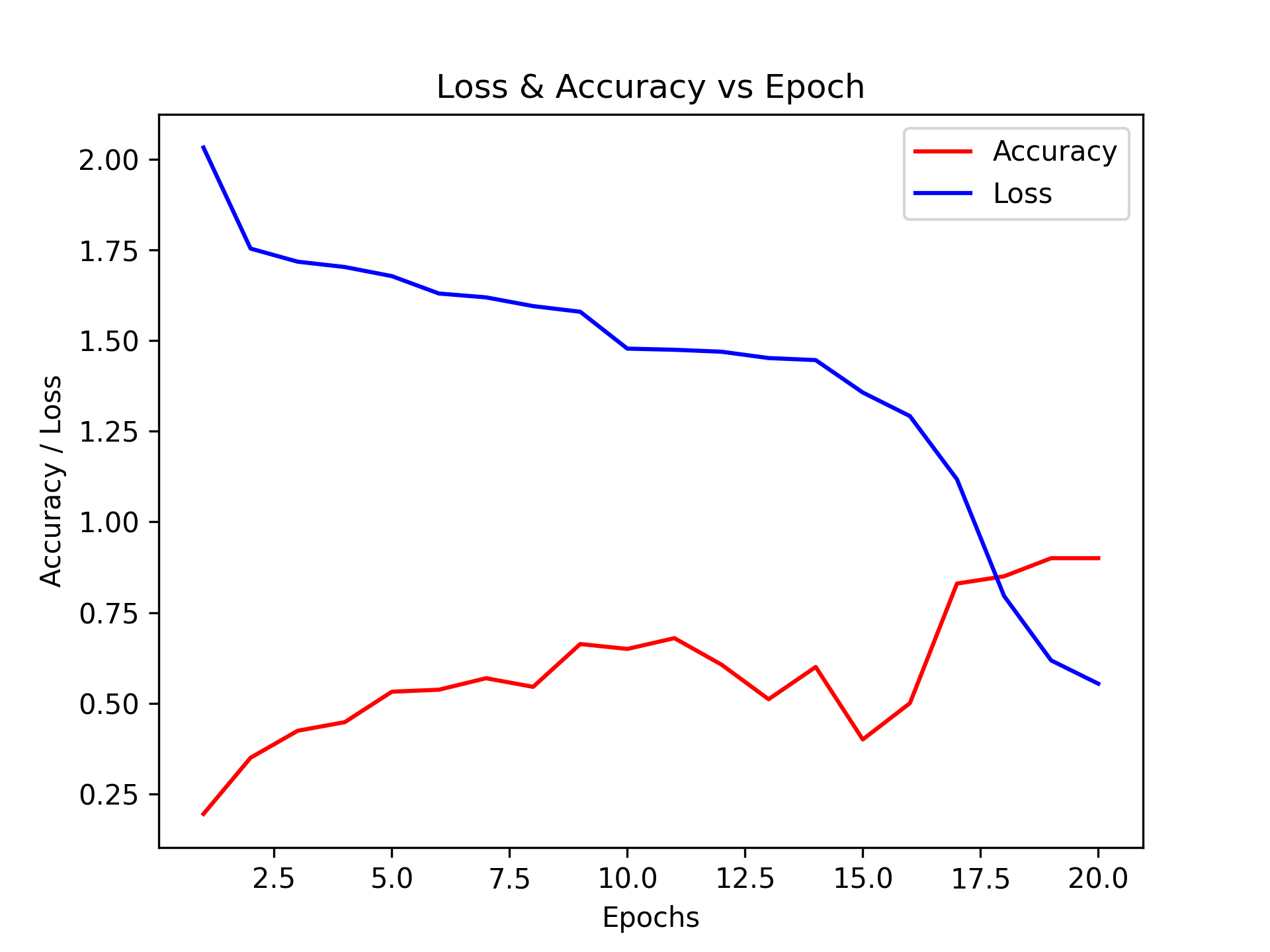}
    \caption{}
    \label{cnnd12x2}
\end{subfigure}\hfill
\begin{subfigure}{\linewidth}
     \includegraphics[width=0.8\linewidth]{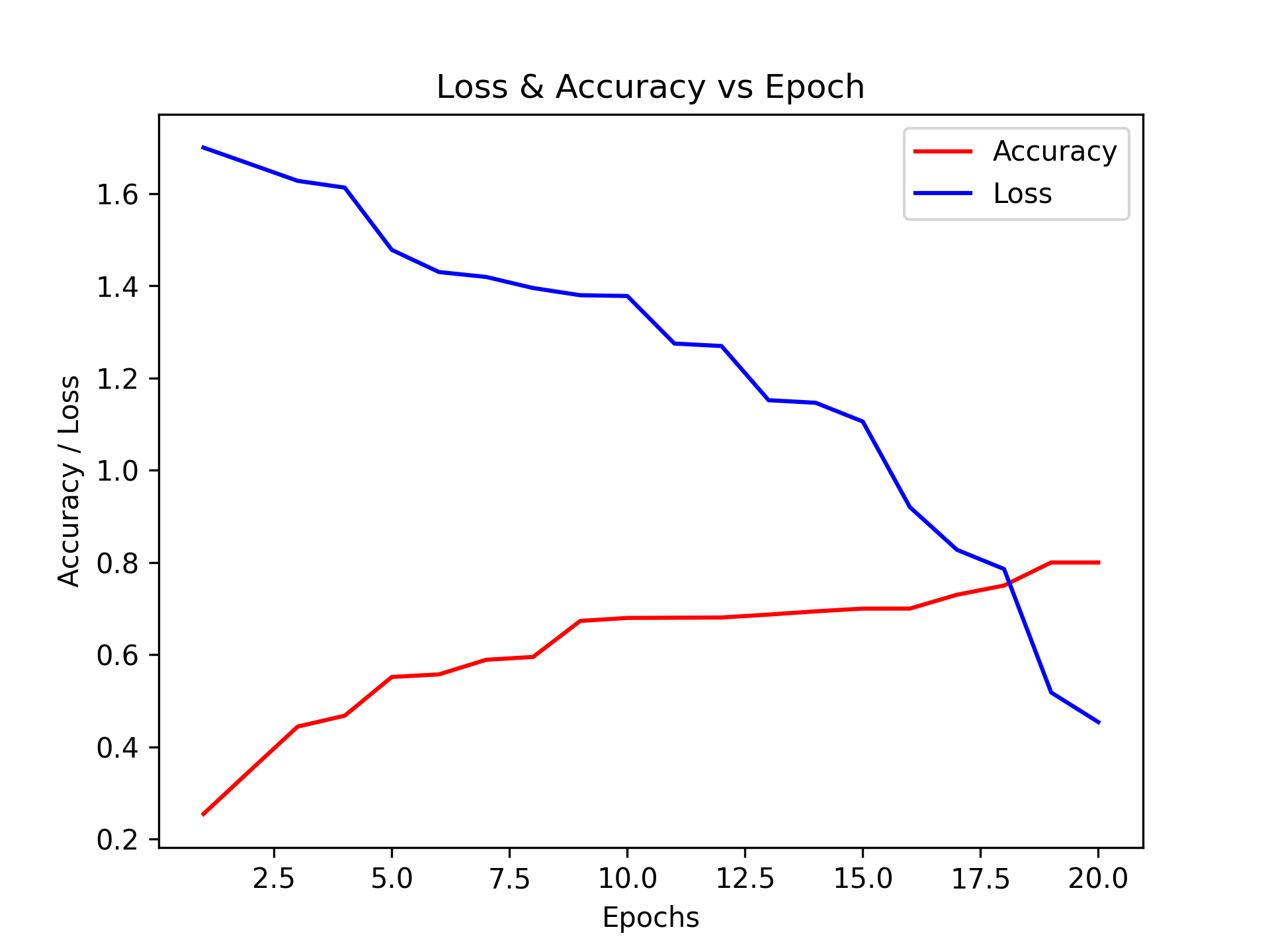}
    \caption{}
    \label{cnnd14x4}
\end{subfigure}\hfill
\begin{subfigure}{\linewidth}
    \includegraphics[width=\linewidth]{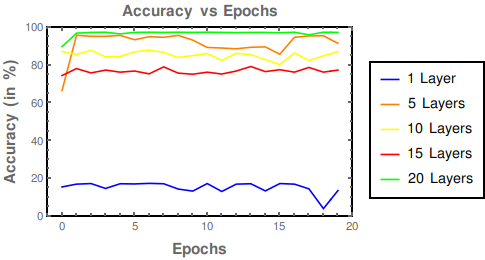}
    \caption{}
    \label{acccarla}
\end{subfigure}\hfill
\begin{subfigure}{\linewidth}
    \includegraphics[width=\linewidth]{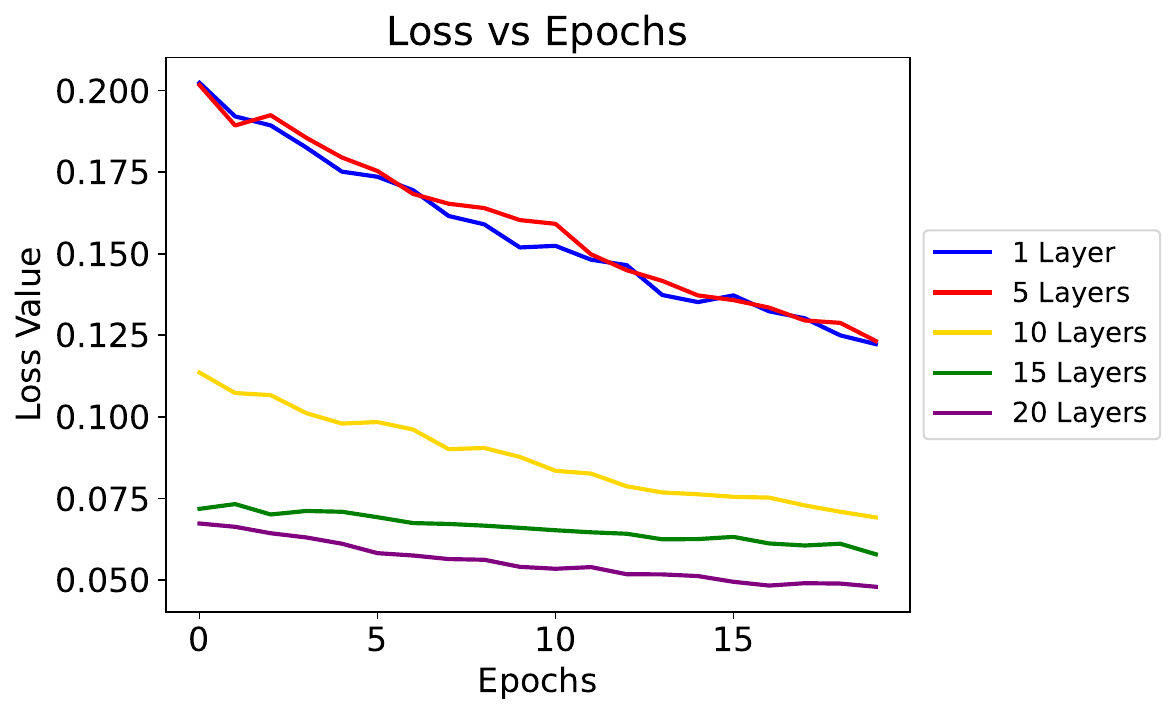}
    \caption{}
    \label{losscarla}
\end{subfigure}\hfill
\caption{
Accuracy and loss results on the Carla dataset: \textbf{(a)} and \textbf{(b)} CNN with 2x2 and 4x4 inputs, and \textbf{(c)} and \textbf{(d)} QNN with 2x2 and 4x4 inputs across varying layers.
}
\label{fig6clas}
\end{figure}

\begin{figure}[htpb]
\begin{subfigure}{\linewidth}
    \includegraphics[width=0.8\linewidth]{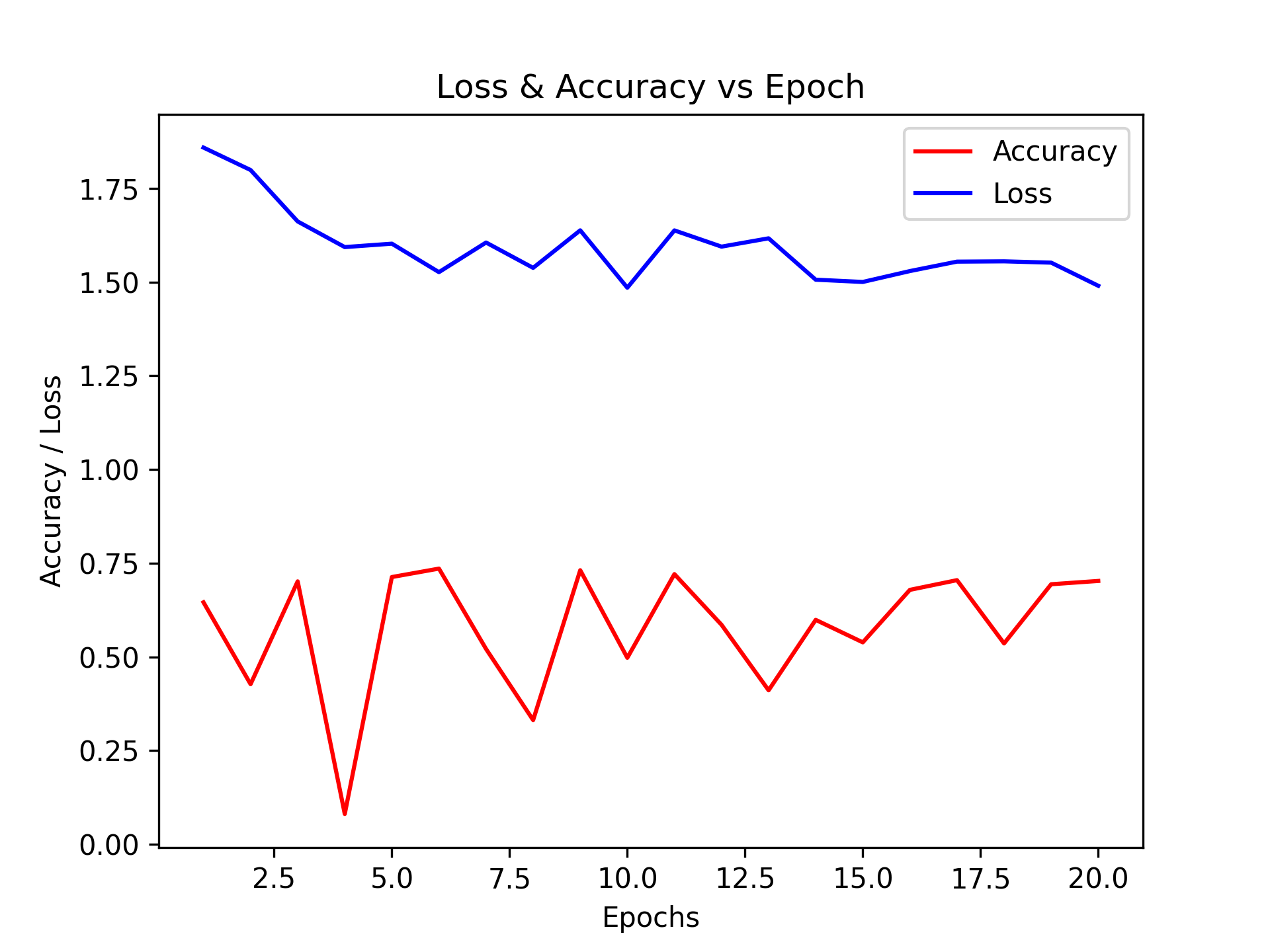}
    \caption{}
    \label{cnn2x2lisa}
\end{subfigure}\hfill
\begin{subfigure}{\linewidth}
     \includegraphics[width=0.8\linewidth]{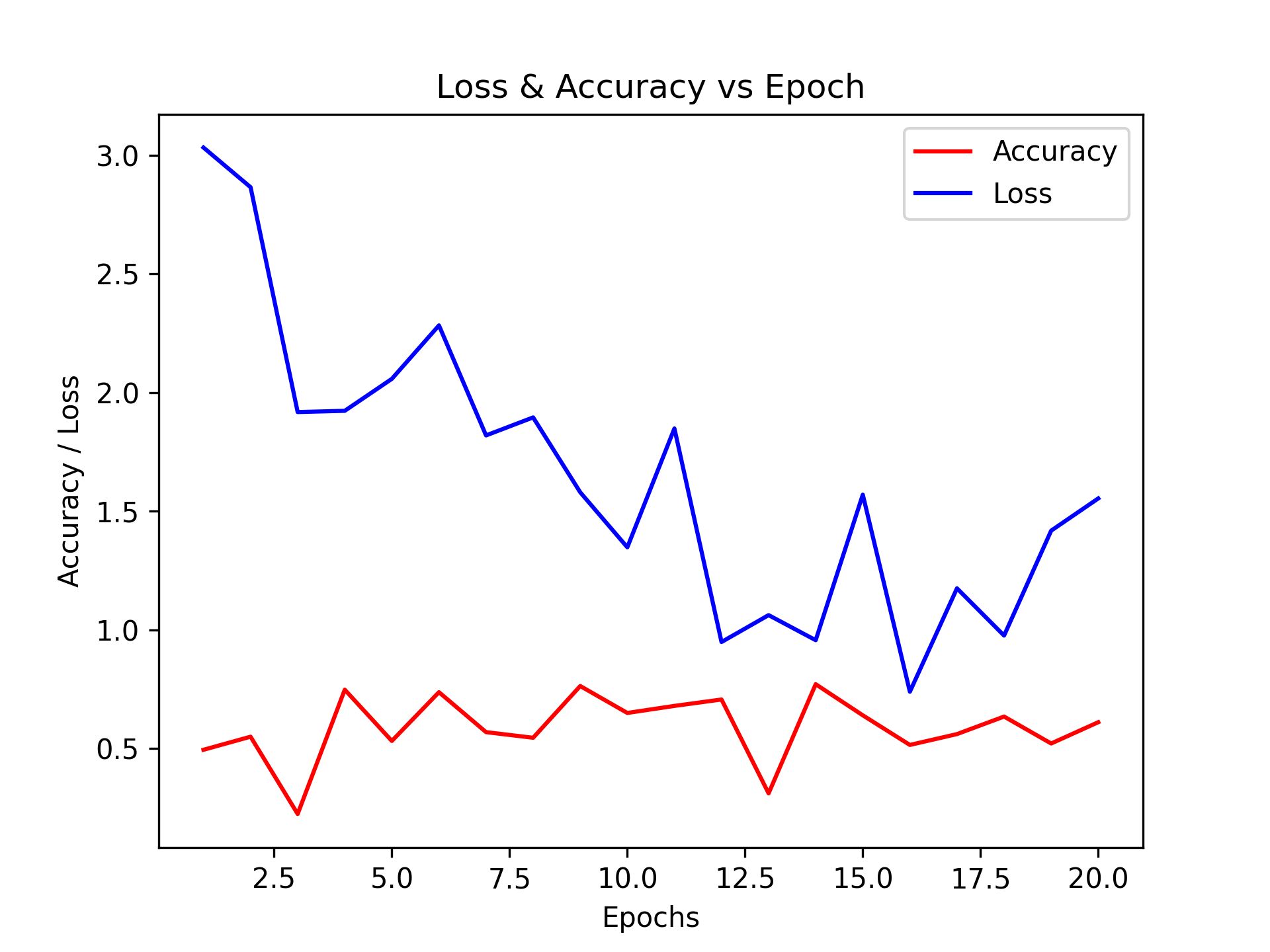}
    \caption{}
    \label{cnn4x4lisa}
\end{subfigure}
\centering
\begin{subfigure}{\linewidth}
    \includegraphics[width=\linewidth]{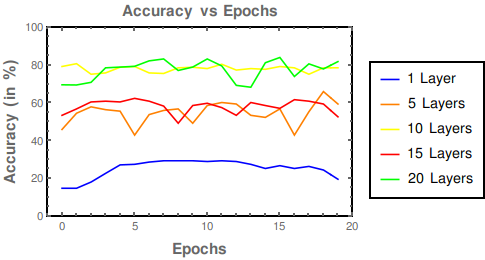}
    \caption{}
    \label{acclisa}
\end{subfigure}\hfill
\begin{subfigure}{\linewidth}
    \includegraphics[width=\linewidth]{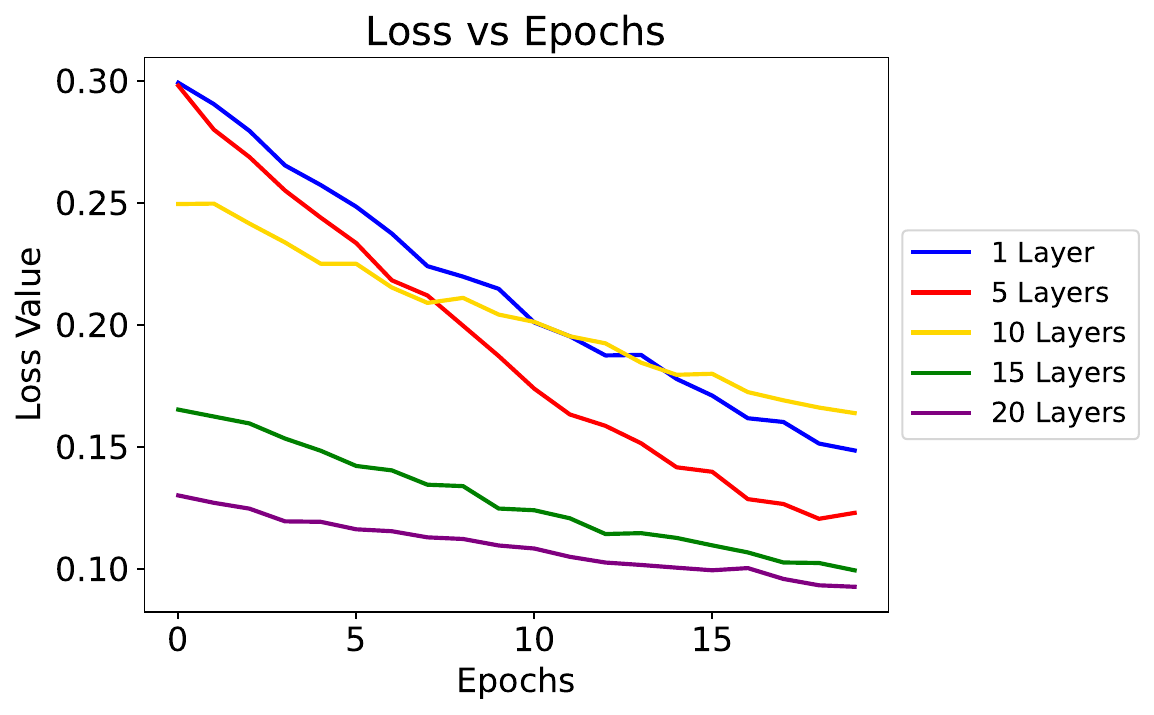}
    \caption{}
    \label{losslisa}
    \end{subfigure}
\caption{
Accuracy and loss results on the cropped Lisa dataset: \textbf{(a)} and \textbf{(b)} CNN with 2x2 and 4x4 inputs, and \textbf{(c)} and \textbf{(d)} QNN with 2x2 and 4x4 inputs across varying layers.
}
\label{lisaresults}
\end{figure}

\subsection{Results\label{sec5}}
\subsubsection{Results of Carla Dataset}
The performance of CNN models with inputs of image sizes of $2\times2$ pixels and $4\times4$ pixels, as shown in Figs. \ref{cnn2x2lisa} and \ref{cnn4x4lisa}, provide insight into the impact of input resolution on model training. The CNN with the $2\times2$ pixel input demonstrates a steady decrease in loss and a corresponding increase in accuracy, with some variability across epochs. 
In contrast, the CNN with the $4\times4$ pixel input exhibits a more tempered reduction in loss and a less pronounced increase in accuracy over the same number of epochs. 
As presented in Fig.~\ref{acccarla} for the QNN, a trend of improving accuracy with the addition of layers up is demonstrated, QNN with one layer shows the least accuracy (around 18\%), which significantly improves as the number of layers is increased to 5 and 10, indicating that a minimal number of layers is insufficient for the QNN to generalize from the dataset effectively. The accuracy reaches a plateau with 20 layers with an accuracy of 100\%.
Additionally, as shown in Fig.~\ref{losscarla}, the QNN presents a clear downward trend across epochs, with the rate of loss reduction being more pronounced in models with more layers. 
As presented in Table \ref{TabI} and discussed above, the models are compared to the other algorithms; in the mean average, the CNN models with $2\times2$ pixel inputs achieved an accuracy of 80\%, while the same model with $4\times4$ pixel inputs remarkably achieved a higher accuracy of 90\%. 
Quantum classification methods using $UU^{\dagger}$ encodings showed varying degrees of success. The $2\times2$ inputs with FRQI encoding reported an accuracy of 66.25\%, which is significantly lower than the classical CNN's performance at the same resolution. 
The variational $UU^{\dagger}$ approach, both with FRQI and NEQR encodings for $2\times2$ inputs, resulted in lower accuracies of 43.21\% and 35.34\%, respectively. These results were mirrored in the $4\times4$ pixel inputs, where the variational $UU^{\dagger}$ with FRQI encoding achieved 43.46\%. The consistency in lower performance across both encoding strategies and resolutions suggests that the variational models may require further refinement to be effective for this particular dataset.
In stark contrast, the QNN with Angle Encoding (AE) and $2\times2$ pixel inputs achieved a strikingly high accuracy of 97.42\%, outperforming all other methods.

\begin{table}[ht]
\caption{Experimental results of all classification methods for Carla dataset. AE: Angle Encoding}
\centering
\begin{tabular}{|c|c|c|c|}
\hline
\textbf{Pixels} & \textbf{Methods} & \textbf{Encodings} & \textbf{Accuracy} ($\%$)\\
\hline
$2\times2$ & CNN & --  & 80\\
$4\times4$ & CNN & --  & 90\\
$2\times2$ & $UU^{\dagger}$  & FRQI & 66.25 \\
$2\times2$   & Variational $UU^{\dagger}$ & FRQI  & 43.21 \\    
$2\times2$ & $UU^{\dagger}$  & NEQR  & 35.34 \\
$2\times2$   & Variational $UU^{\dagger}$ & NEQR  &35.34 \\  
$4\times4$ & $UU^{\dagger}$ & FRQI & 72.44 \\
$4\times4$    & Variational $UU^{\dagger}$ & FRQI  & 43.46 \\        
$2\times2$ & QNN  & AE &97.42 \\
\hline
\end{tabular}

\label{TabI}
\end{table}
\subsubsection{Results of Cropped Lisa Dataset}
In Fig. \ref{cnn2x2lisa}, the CNN with $2\times2$ pixel inputs displays a fluctuating trend in accuracy, indicating potential challenges in the model's learning stability. Despite these fluctuations, there is a general upward trend in accuracy, suggesting that the model is improving its predictive capabilities as training progresses. The loss for this model exhibits less variability, showing a gentle, albeit inconsistent, decrease. This pattern of loss reduction hints at the model's capacity to gradually minimize the prediction error over time, although the fluctuations in accuracy suggest that the model might benefit from further tuning to enhance its generalization ability.
Fig. \ref{cnn4x4lisa} portrays the CNN with $4\times4$ pixel inputs, where both loss and accuracy exhibit significant volatility across epochs. The accuracy fluctuates widely without a clear upward trend, which could be indicative of a model struggling to fit the data effectively. The loss initially decreases but then follows a variable pattern, with moments where it increases, suggesting that the model may be encountering difficulties in optimizing its parameters. 
The behavior of QNN models with varying numbers of layers, as presented in Fig. \ref{losslisa}, indicates a clear trend where models with an intermediate number of layers (specifically, ten and fifteen) show a more marked reduction in loss and an increase in accuracy, as shown in Fig. \ref{acclisa}. 
Additionally, as presented in Table \ref{TabII}, classical CNN models with $2\times2$ and $4\times4$ pixel inputs report accuracy levels of 77.09\% and 73.13\%, respectively. The superior performance of the $2\times2$ model may be attributed to a more effective feature capture that this scale affords or possibly a more favorable alignment with the network's architecture for this particular dataset, as illustrated in the loss and accuracy plots.
And for quantum methods, the use of $UU^{\dagger}$ encodings with FRQI on $4\times4$ inputs results in an accuracy comparable to classical CNNs, specifically 76.95\%, indicating that quantum methods hold their own against classical counterparts when appropriate encoding is applied. However, the variational $UU^{\dagger}$ models exhibit a notable drop in accuracy, achieving only 43.24\% for $2\times2$ inputs and a similar 43.33\% for $4\times4$ inputs. This decline may underscore the challenges inherent in optimizing variational quantum circuits.
The implementation of NEQR encoding with $UU^{\dagger}$ produces a further decrease in accuracy to 35.34\%, irrespective of the approach being variational or not, for the $2\times2$ inputs. This outcome suggests that NEQR encoding may be less effective for this dataset within the tested frameworks.
The standout result is the QNN with Angle Encoding (AE) for $2\times2$ pixel inputs, which achieves the highest recorded accuracy of 84.08\%. 

\begin{table}[ht]
\caption{Experimental results of all methods of classification for the Cropped Lisa dataset. AE: Angle Encoding}
\centering
\begin{tabular}{|c|c|c|c|}
\hline
\textbf{Pixels} & \textbf{Methods} & \textbf{Encodings} & \textbf{Accuracy} ($\%$)\\
\hline
$2\times2$ & CNN & --  & 77.09\\
$4\times4$ & CNN & --  & 73.13\\
$2\times2$ & $UU^{\dagger}$  & FRQI & 72.82 \\
$2\times2$   & Variational $UU^{\dagger}$ & FRQI  & 43.24 \\    
$2\times2$ & $UU^{\dagger}$  & NEQR  & 35.34 \\
$2\times2$  & Variational $UU^{\dagger}$ & NEQR  &35.34 \\  
$4\times4$ & $UU^{\dagger}$ & FRQI & 76.95 \\
$4\times4$ & Variational $UU^{\dagger}$ & FRQI  & 43.33 \\    $2\times2$ & QNN  & AE & 84.08 \\
\hline
\end{tabular}

\label{TabII}
\end{table}
\begin{figure}[htpb]
    \centering
\begin{subfigure}{\linewidth}
        \centering
        \includegraphics[width=0.8\linewidth]{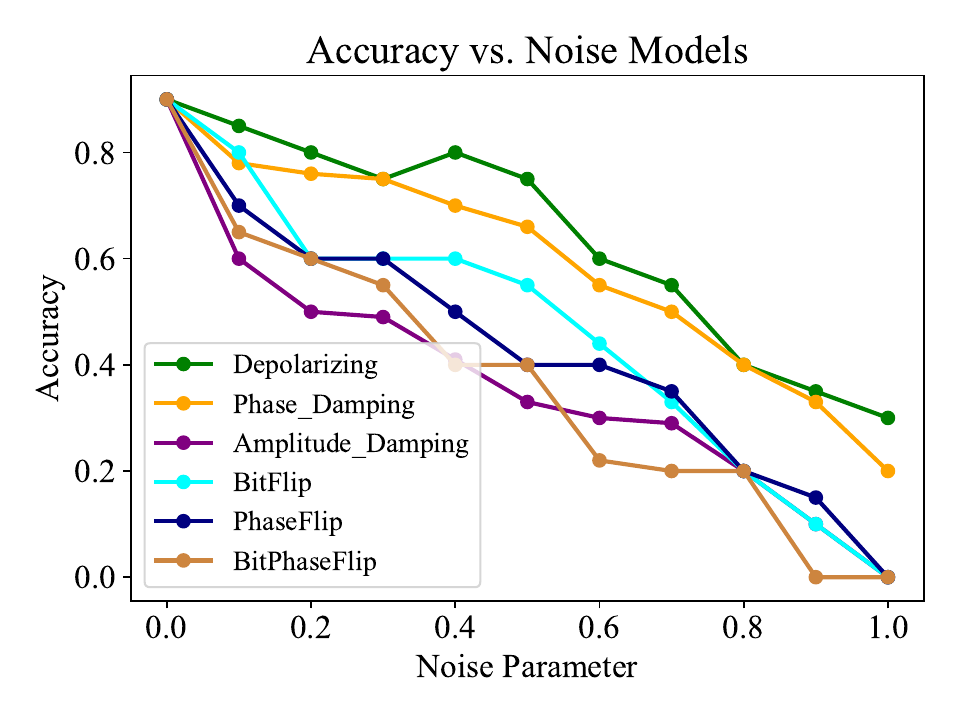}
        \caption{}
        \label{noise1}
    \end{subfigure}
\begin{subfigure}{\linewidth}
        \centering
        \includegraphics[width=0.8\linewidth]{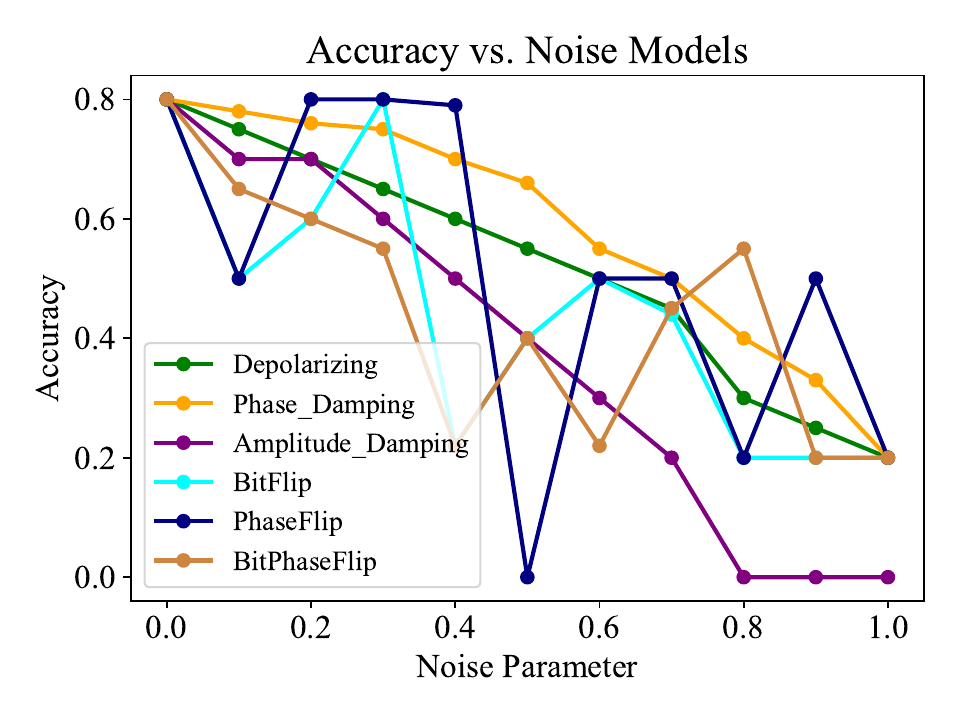}
        \caption{}
        \label{noise2}
    \end{subfigure}    
    \caption{Representation of QNN accuracy across diverse noise models on (a) Carla and (b) Cropped Lisa.}
    \label{noise}
\end{figure}

\subsection{Noisy Simulation}
In evaluating the robustness of the QNN model against six distinct noise models, the findings exhibit significant variations in performance across different types of noise, as shown in Fig. \ref{noise}. 
The model exhibits a steady decline in accuracy for the Carla dataset with increasing depolarizing noise, suggesting a stable but diminishing trend. This can be attributed to how depolarizing noise introduces random Pauli errors between the single-qubit and entangling gates, which generates a gradual loss of coherence in the circuit. As more noise increases, more gates will be impacted; therefore, there is a progressive decay in accuracy. Conversely, phase damping noise demonstrates remarkable stability, with the model maintaining high accuracy even at elevated noise intensities. This type of noise acts essentially only on the phase of the qubits, while not affecting the qubit population; hence, this explains the robustness of such a model. All the gates that are not so dependent on phase coherence, such as rotation around the X and Y axes, are less affected; therefore, more stable performances are achieved. Notably, amplitude damping noise markedly affects performance, with accuracy dropping to zero as the noise parameter surpasses 0.7, which leads to energy loss from excited states to ground states, fundamentally disrupting quantum gates that rely on superpositions. As qubits reset to their ground state, the model loses critical information, causing a steep drop in accuracy. Both bitflip and phaseflip noise models elicit an irregular accuracy pattern, reflecting the QNN model's non-linear and complex response to these noises. Specifically, the model under bitflip noise fluctuates in performance, and under phaseflip noise, it shows a significant drop in accuracy at a 0.5 noise parameter, followed by a subsequent recovery. This irregularity is due to how bitflip noise randomly flips qubits, disrupting operations that rely on the correct qubit state being preserved. Similarly, phaseflip noise affects qubit phases, leading to sudden drops in accuracy when phase-sensitive gates are disrupted. The bitphaseflip noise consistently leads to a decline in accuracy, though with some fluctuations.

Subsequent analysis with the cropped Lisa dataset offers additional perspectives. The model demonstrates higher overall accuracy across all noise types than the Carla dataset. This can be attributed to the differences between the datasets, where cropped Lisa images may contain more distinguishable features, making the quantum gates less sensitive to errors introduced by noise. The depolarizing noise model shows high resilience, maintaining accuracy above 0.55 even at maximum noise levels. While still robust, phase damping noise exhibits a more noticeable decline in accuracy compared to the Carla dataset. This trend is likely due to phase-sensitive gates playing a larger role in extracting features from this dataset. Like the Carla dataset, amplitude damping noise significantly compromises accuracy, particularly beyond a noise parameter of 0.4. Bitflip and phase flip models in the cropped Lisa dataset suggest a more linear decrease in accuracy with increasing noise levels, unlike the varied response seen in the Carla dataset. This behavior suggests a more predictable impact of these noises, likely due to better correlation between the quantum states and the classification task. In this context, the bitphase flip model displays a moderate yet more pronounced decline in accuracy as noise intensifies.

\subsection{Discussion}
For the Carla dataset, it is observed that the classical CNNs benefit from higher input resolutions, with the $4\times4$ pixel input model achieving a 90\% accuracy rate, demonstrating the advantage of capturing more detailed features. However, the quantum approaches, particularly the QNN with AE) and $2\times2$ pixel inputs, excel with an impressive accuracy of 97.42\%, surpassing the classical models. This superiority suggests that QNNs when optimized with suitable encodings, could potentially redefine the benchmarks for traffic image classification.
The analysis is further deepened with the Cropped Lisa dataset, where CNNs present variability in learning, indicating a potential for enhancement through model refinement. In contrast, QNNs show a pronounced increase in performance with an intermediate number of layers yet also hint at a performance ceiling beyond which additional layers do not correlate with improved accuracy.
While certain types of noise, like phase damping, are well-tolerated, others, such as amplitude damping, significantly impair the model's performance. 
The results' comparison with existing methodologies, including classical and quantum models, as presented in Table \ref{TabIII}, places the QNN model's achievements in a broader context, showing a clear advancement over the current benchmarks. 
\begin{table}[ht]
\caption{Comparative accuracy results for traffic sign detection across various existing studies. FD: Fusion Detection; TL: Transfer Learning.}
\centering
\begin{tabular}{|c|c|c|}
\hline
\textbf{Models} & \textbf{Accuracy ($\%$)} \\
\hline
IARA-TLD \cite{possatti2019traffic} & 69.53\\
Faster R-CNN \cite{bach2018deep}  & 71.2\\
FD \cite{li2017traffic}  & 71.50 \\
R-CNN \cite{app112210713}  & 78.4 \\
QNN \cite{kuros2022traffic} & 94.40 \\
Quantum TL \cite{khatun2024quantum} &94.62\\
Anchor-Free Detector \cite{zhang2024robust} & 93.95\\ 
YOLOv5 \cite{li2023improved} &   96.1\\
LE-CNN \cite{khalifa2024real} & 96.5\\
\textbf{Our model} & \textbf{97.42}\\
\hline
\end{tabular}

\label{TabIII}
\end{table}

However, in realistic traffic systems, several challenges still exist with respect to the implementation of quantum models. One key issue is that quantum systems need to be integrated into existing classical infrastructures since most of the management systems in traffic signals do not have the capability to handle QC. Therefore, compatibility issues between classical and quantum systems are very important. Additionally, latency is crucial since quantum computations may introduce delays that affect real-time decisions. Therefore, hardware advancements and algorithm design must address these timing constraints to ensure that the system continues operating efficiently within the required timeframes. In addition, traffic systems deal with very sensitive information, and data privacy remains one of the big challenges. Quantum-enhanced systems should be designed to integrate robust privacy-preserving techniques, including quantum encryption and secure communications, towards protecting data while considering the regulations regarding privacy. Overcoming these challenges is the prerequisite for successful model deployment through quantum models in practical traffic applications.
\section{Conclusion \label{sec6}}
This paper introduces a groundbreaking exploration into using QNNs to advance VRCS. By harnessing the capabilities of quantum methods, including $UU^{\dagger}$, variational $UU^{\dagger}$ algorithms, and, most notably, QNNs, the research addresses the complex task of classifying traffic light images into red, green, or yellow categories. This classification is essential for improving traffic management decisions.
The investigation reveals that while the $UU^{\dagger}$ and variational $UU^{\dagger}$ methods, alongside FRQI and NEQR encoding techniques, establish a solid base for quantum-based image classification, it is the QNN model that marks a significant advancement in the field. The QNN algorithm stands out by achieving remarkable accuracies of 97.42\% and 84.08\% on two distinct datasets. This performance, achieved with minimal training data and reduced image dimensions, highlights the superior generalization capability of QNNs.
The paper underscores the innovative role of QNNs in ITS, presenting a significant step toward embedding quantum-augmented intelligence in vehicular technologies. It details the QNN model's robustness against diverse noise types, showcasing its applicability in real-world scenarios. Despite its varied sensitivity to noises such as depolarizing, amplitude damping, bitflip, phaseflip, and bitphaseflip, the model's resilience, particularly to phase damping noise, exemplifies its potential for deployment in ITS.
For future direction, it is important to analyze and study the scalability of quantum circuits for practical use in real systems, focusing on the number of qubits and the complexity of the gates; understanding this aspect will help us understand the scalability of the circuits and also of their hardware efficiency, allowing their effective realization on current and future quantum technologies toward practical applications.
The contribution of this research to the field is profound, suggesting a shift in how QC can be applied to traffic image classification and ITS. The QNN model, with its high accuracy and demonstrated resilience to noise, sets a new standard for vehicle road cooperation systems, embodying a transformative approach to the VRCS.
In conclusion, the findings of this study not only validate the efficacy of QNNs over existing classical and quantum models but also highlight the quantum leap in machine learning applications for ITS and 6G. 

\section*{Acknowledgment}
Work by N.I was supported in part by the NYUAD Center for Quantum and Topological Systems (CQTS), funded by Tamkeen under the NYUAD Research Institute grant CG008.

\ifCLASSOPTIONcaptionsoff
\newpage
\fi

\bibliographystyle{IEEEtran}

\bibliography{IEEE}

\vfill

\end{document}